\documentclass[conference,compsoc]{IEEEtran}
\usepackage{amsmath,amssymb,amsfonts}
\usepackage{float}
\usepackage{graphicx} 
\usepackage{amsmath} 
\usepackage{listings}
\usepackage{multirow}
\usepackage{pifont}
\usepackage{ragged2e}
\usepackage{subcaption}
\usepackage{url}
\usepackage{xcolor}

\newcommand{\xmark}{\ding{55}}

\lstdefinestyle{fstar}{
    language=Haskell, 
    basicstyle=\ttfamily\footnotesize,
    keywordstyle=\color{blue}\bfseries,
    commentstyle=\color{green},
    stringstyle=\color{red},
    numberstyle=\tiny\color{gray},
    numbers=left,
    stepnumber=1,
    numbersep=5pt,
    backgroundcolor=\color{white},
    showspaces=false,
    showstringspaces=false,
    showtabs=false,
    frame=single,
    rulecolor=\color{black},
    breaklines=true,
    breakatwhitespace=false,
    escapechar=~,
    morekeywords={interface, struct, enum},
    captionpos=b
}

\def\BibTeX{{\rm B\kern-.05em{\sc i\kern-.025em b}\kern-.08em
    T\kern-.1667em\lower.7ex\hbox{E}\kern-.125emX}}

\makeatletter 
\newcommand{\linebreakand}{%
  \end{@IEEEauthorhalign}
  \hfill\mbox{}\par
  \mbox{}\hfill\begin{@IEEEauthorhalign}
}
\makeatother

\begin{document}

\title{Towards Provable Security in Industrial Control Systems Via Dynamic Protocol Attestation}

\author{\IEEEauthorblockN{Arthur Amorim}
\IEEEauthorblockA{\textit{University of Central Florida}\\
Florida, USA \\
arthur.amorim@ucf.edu}
\and
\IEEEauthorblockN{Trevor Kann}
\IEEEauthorblockA{\textit{Carnegie Mellon University}\\
Pittsburgh, USA \\
tkann@cmu.edu} 
\and
\IEEEauthorblockN{Max Taylor}
\IEEEauthorblockA{\textit{Idaho National Laboratory}\\
Idaho Falls, USA \\
maxwell.taylor@inl.gov}
\and
\IEEEauthorblockN{Lance Joneckis}
\IEEEauthorblockA{\textit{Idaho National Laboratory}\\
Idaho Falls, USA \\
lance.joneckis@inl.gov}
}

\maketitle

\begin{abstract}
Industrial control systems (ICSs) increasingly rely on digital technologies vulnerable to cyber attacks. Cyber attackers can infiltrate ICSs and execute malicious actions. Individually, each action seems innocuous. But taken together, they cause the system to enter an unsafe state. These attacks have resulted in dramatic consequences such as physical damage, economic loss, and environmental catastrophes. This paper introduces a methodology that restricts actions using protocols. These protocols only allow safe actions to execute.  Protocols are written in a domain specific language we have embedded in an interactive theorem prover (ITP). The ITP enables formal, machine-checked proofs to ensure protocols maintain safety properties. We use dynamic attestation to ensure ICSs conform to their protocol even if an adversary compromises a component. Since protocol conformance prevents unsafe actions, the previously mentioned cyber attacks become impossible. We demonstrate the effectiveness of our methodology using an example from the Fischertechnik Industry 4.0 platform. We measure dynamic attestation's impact on latency and throughput. Our approach is a starting point for studying how to combine formal methods and protocol design to thwart attacks intended to cripple ICSs.
\end{abstract}

\begin{IEEEkeywords}
Formal methods, Correctness proofs, Industrial automation, Industrial control, Software and System Safety.
\end{IEEEkeywords}

\section{Introduction}

Industrial control systems (ICSs) are large-scale, distributed, cyber-physical systems controlled by software. ICSs are increasingly targeted in cyber attacks.  For example, the Stuxnet worm used malicious commands to destroy centrifuges \cite{stuxnet}. In 2014, a similar attack caused immense damage to a blast furnace in a steel mill \cite{steelMill}. Blast furnaces must be shut down according to a specific protocol or risk damage. Attackers infiltrated the steel mill's ICSs and violated the shutdown protocol.

\begin{figure}
    \centering
    \includegraphics[scale=.75]{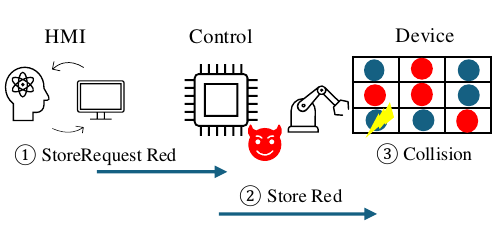}
    \caption{Example attack. An adversary infiltrates the control unit to cause a collision in the warehouse.}
    \label{fig:intro-example-attack}
\end{figure}

Figure \ref{fig:intro-example-attack} shows an attack against an ICS controlling a high bay warehouse storing red, white, and blue items. This ICS contains a human machine interface (HMI), a control unit, and low-level device drivers. These components interact by executing remote procedure calls (RPCs). At the start of this attack, the warehouse is full. Then, the operator executes a \textbf{StoreRequest} RPC. Alone, this RPC is benign: the controller is under no obligation to store an item in a full warehouse. But in this controller resides an attacker who previously infiltrated the ICS. The attacker seizes their opportunity to execute a \textbf{Store} RPC on the hardware drivers. Since there is no empty bay available, a collision occurs.

An ICS's software components can become compromised at any moment. Compromises are increasing with the rise of ``Industry 4.0.'' Industry 4.0 integrates manufacturing and digital technology \cite{ibm_industry_4_0}.  Often, digital technology uses the public internet. Presence on the internet exposes new and existing vulnerabilities to attackers. Attackers exploit vulnerabilities to compromise ICS components. Compromised components can execute arbitrary RPCs. Hence, we assume an attacker can execute arbitrary, malicious RPCs. Malicious RPCs can lead to unsafe behavior, as in Figure \ref{fig:intro-example-attack}.


It is possible, albeit challenging, to show an ICS behaves safely using formal methods. Formal methods are a family of mathematical techniques for reasoning about systems. Formal methods have two popular techniques: model checking and theorem proving. Model checking has been applied in previous research efforts to obtain safety poofs for ICSs~\cite{hodaOntology,modelCheckingPlcFbds,  translationBasedModelChecking, plcVerif, applyingModelChecking}. However, model checking suffers from the state explosion problem. This problem prevents model checking from scaling to larger ICSs, as shown in \cite{applyingModelChecking}. 

Scaling safety proofs to larger ICSs is possible using theorem proving. However, theorem proving notoriously requires significant effort. For example, it cost 20 person-years to verify seL4, an OS written in $\approx 8000$ lines of C~\cite{sel4Paper}. But this effort was only necessary because the totality of seL4 was verified. All other factors remaining equal, verifying less software requires less effort.

Our main idea is to restrict sequences of RPCs to a protocol. We prove protocol conformance implies the ICS behaves safely. Thus, unsafe behavior can only ever occur as a result of executing a non-conformant RPC. We keep non-conformant RPCs from executing using a dynamic protocol attestation system. When the attestation system encounters a non-conformant RPC, it triggers a fail-safe. Fail-safes are always safe to execute. 

In our approach, engineers describe their system's RPCs using an interactive theorem prover (ITP). ITPs are programs that help engineers write formal, machine-checked proofs. This paper uses F* \cite{fstar}, a state-of-the-art ITP with sophisticated proof automation. Inside F*, we have embedded a domain specific language (DSL). Our DSL helps engineers design RPC protocols. Engineers use F*'s proof automation to prove that protocol conformance implies safety.

We provide automation to generate code for RPC frameworks from our DSL.  Although ITPs like F* can help verify designs, they are not suitable for writing ICS software. ICS software is often written in embedded languages like C or C++. Embedded languages are usually not memory or type safe. This lack of safety makes RPC implementations more vulnerable to attacks. We can reduce the risks posed by this attack vector by leveraging off-the-shelf RPC frameworks. RPC frameworks use high-level message specifications to generate "boilerplate" code. With boilerplate code automatically generated, software engineers can concentrate on essential application logic. Our automation extracts message specifications for RPC frameworks directly from an engineer's F* specification. 

Critically,  our approach makes formally verifying all software in an ICS unnecessary. Instead, we only need to verify software that interfaces with physical mechanisms. For example, the low-level software that drives a servo or reads a thermometer. It is easy to identify these software components. In our formal model, they directly interact with the system state. Our proofs show that the ICS behaves safely if RPC semantics are accurately modeled; i.e., if the software implementation matches the F* semantics.

To summarize our contributions, we:
\begin{itemize}
\item \textbf{Propose a new methodology for developing ICS software.} Our methodology enables us to construct new communication protocols in F*, a state-of-the-art ITP. We use F* to prove that protocols maintain critical safety invariants.
\item \textbf{Provide automation to extract RPC framework inputs from F*.} This generates the code necessary to create implementations of the protocol.
\item \textbf{Show how to use dynamic protocol attestation to reduce the verification burden.} 
\item \textbf{Provide experimental results that highlight the potential of our approach.} Dynamic attestation checking adds moderate overhead. In many applications, the overhead will be acceptable. Future work will consider how to further reduce the overhead.
\end{itemize}

\section{Background}

\subsection{Fischertechnik}

\subsubsection{Overview}
Industry and academia use Fischertechnik's embedded platforms~\cite{FT} to address the challenges posed by Industry 4.0~\cite{ibm_industry_4_0}.  Industry 4.0 increases the complexity and concurrency of ICS software. These increases have lead to vulnerabilities not previously found in embedded systems. Members of industry and academia are mitigating these vulnerabilities with new technical solutions, often validated on Fischertechnik's platforms.



The Fischertechnik was initially used in studies involving controller verification with model checking \cite{Cattel1996,Cattel1995,Gourcuff2008,Cattel1995}, in an example implementation of model-driven systems engineering \cite{model1,model2}, and in a program analysis-based PLC code vetting study \cite{Zhang2022}. 

Researchers have also used Fischertechnik platforms to study ICS security. For example, researchers used Fischertechnik platforms to study runtime detection of improper parameters in ICSs \cite{Zhang2022}. They were also used to study decentralized monitoring of linear temporal logic (LTL) formulae \cite{Omar2021}.  Finally, a contract-based hierarchical resilience management framework which improves resilience when dealing with component failures used a Fischertechnik platform to perform its evaluation \cite{Haque2018}.

\subsubsection{High Bay Warehouse}
\begin{figure}[H]
    \centering
    \includegraphics[width=0.65\linewidth]{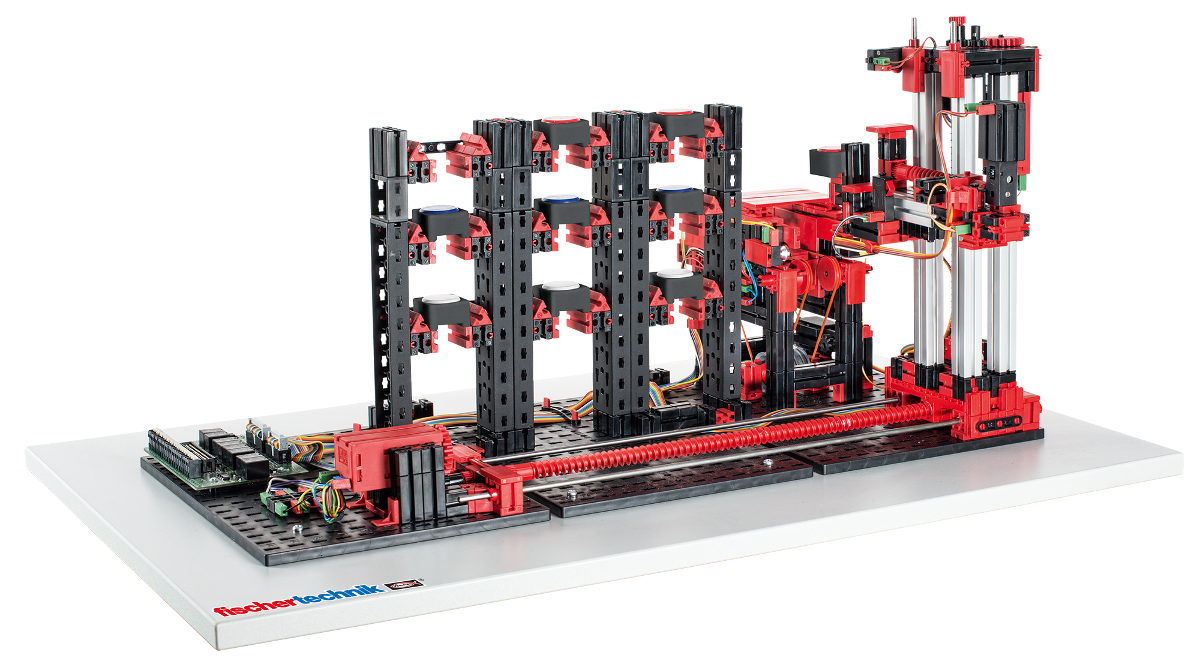}
    \caption{Image of the Fischertechnik's High Bay Warehouse from \cite{fischertechnikHBW}}
    \label{fig:ft-hbw}
\end{figure}

The Fischertechnik factory stores manufactured items in its high bay warehouse (HBW). The HBW is shown in Figure \ref{fig:ft-hbw}. The HBW provides an interface that displays items to customers. The HBW consists of nine bays that can store at most one item. The HBW interacts closely with the vacuum gripper. The vacuum gripper delivers and retrieves items from storage. The HBW also has an arm that moves items into and out of bays. We chose to implement our approach on the HBW due to its intuitive functionality and straightforward interactions with other components.

\subsubsection{Critical Safety Properties}
\label{sec:safety-properties}

The high bay warehouse is vulnerable to two dangerous scenarios:
\begin{enumerate}
\item Storing a new item when the high bay warehouse is already full.
\item Attempting to retrieve an item that is not stored in a bay.
\end{enumerate}

Therefore, an ICS's software implementation must enforce three invariants. First, that the HBW is not full before it attempts to store an item. Second, that the HBW only processes a customer's order if it contains the desired item. Third, that the state of the bays is accurately tracked. In this example the safety property is that none of these invariants are violated.

\subsection{Attack Model}

\begin{figure}
    \centering
    \includegraphics[width=.8\linewidth]{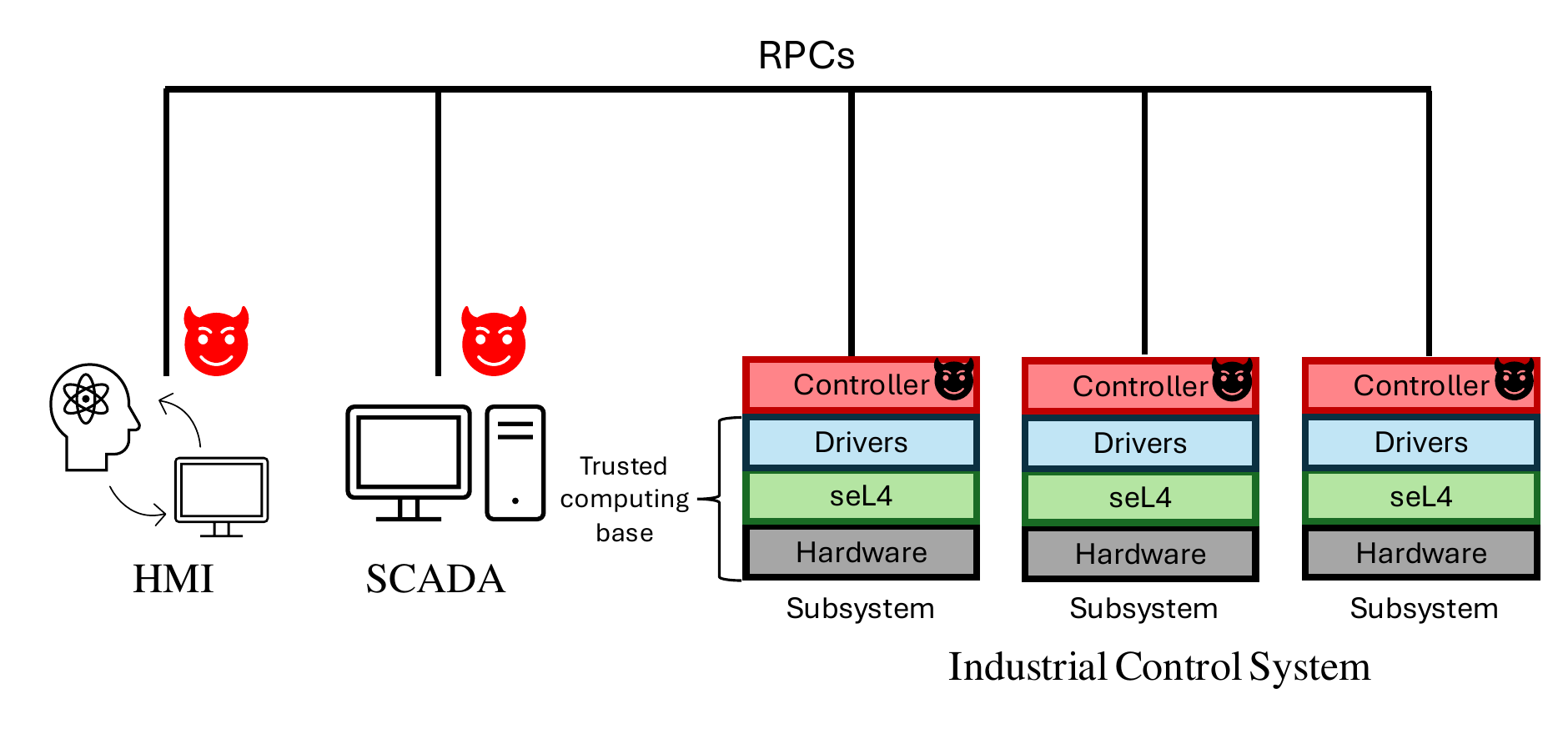}
    \caption{Overview of an ICS. The demon icon is located adjacent to components that we assume an attacker controls.}
    \label{fig:attack vector}
\end{figure}

\subsubsection{Threats}
We consider a threat model where adversaries seek to compromise ICS controllers. ICS controllers use low-level hardware drivers to manipulate the physical world. ICS controllers support emerging industry 4.0 trends by interfacing with the public internet. Interfacing with the internet jeopardizes the security of an ICS controller. Therefore, we assume an ICS controller may become compromised at any moment. Compromised ICS controllers may execute RPCs that place the ICS at-risk. 

\subsubsection{Attacker}
We assume the attacker is able to execute arbitrary RPCs on any controller in the ICS. Figure \ref{fig:attack vector} shows possible attack vectors. ICS controllers communicate with low-level device drivers. Thus, the attacker is able to execute malicious actions from compromised controllers. The attacker may also execute malicious RPCs on uncompromised controllers connect by a network. By trusting input from malicious RPCs, uncompromised controllers can also take a system to an unsafe state.

\subsubsection{Trusted Computing Base}
We assume attackers cannot control the operating system, low-level hardware drivers, or hardware. High assurance operating systems like seL4 \cite{sel4} have formally verified security properties. These properties enable strong isolation of software components. We assume that engineers have verified the behavior of hardware drivers. We trust that hardware functions as intended. We also assume attackers have not compromised the RPC framework. We further assume the RPC framework always executes RPCs. Finally, we assume attackers have not compromised our dynamic protocol checker.

\subsection{Interactive Theorem Provers}
Interactive theorem provers are tools that help engineers construct computer-verified mathematical proofs. Engineers write proofs in a language the ITP provides. Often, ITPs provide purely functional languages based on the Calculus of Inductive Constructions \cite{cic}. The Calculus of Inductive Constructions (CIC) is a type theory for expressing proofs. Proofs in CIC are constructed following the Curry-Howard isomorphism \cite{curryHoward}. The Curry-Howard isomorphism treats types as propositions and values as proofs. Writing a proof of a proposition amounts to writing a term whose type corresponds to the proposition. Type-checking verifies that the proof is free from mistakes in reasoning. 

ITPs provide automation that helps users write terms. Writing a term that inhabits a type is often awkward. As the type's corresponding proposition grows more complex, the situation becomes worse. To help remedy the situation, ITPs often provide tactic systems. Tactics are examples of metaprograms: programs that write programs. Tactics construct terms by exposing commands to engineers. These commands more closely match how humans think about proofs. Commands need not even be correct because the term a command produces is ultimately type-checked. Users can extend ITPs with their own tactics without compromising the soundness of their proofs.

 F* \cite{fstar} is a new ITP that makes it even easier for engineers to write proofs. Tactic systems require users to explicitly justify each step in their reasoning. But many steps in reasoning are obvious: equality is symmetric, addition is commutative, etc. Often times, humans make these steps without ever saying so. F* mimics human reasoning by integrating the Z3 SMT solver with its tactic system \cite{z3}. SMT solvers are tools that automatically prove propositions. Since it is impossible to automatically prove all propositions, SMT solvers may fail. When Z3 succeeds, an engineer is relieved of the burden. When Z3 fails, the engineer can use a combination of programming and tactics to write the proof. Z3 still fills gaps between steps, thus tremendously reducing the proof burden. 

\section{Methodology}
This section describes how to construct protocol specifications and prove that safety properties hold on the protocol. Communication protocols describe legal sequences of RPCs. Ultimately, a communication protocol is a formal language inhabited by sequences of RPCs called \emph{traces}. RPCs have semantics that our methodology formally models in F*.  In the semantics, an RPC that causes the ICS to behave unsafely causes the state to become \textbf{Wrong}.  We embed a DSL for describing communication protocols in F*. Using F*, we can verify conformance to a protocol implies the ICS does not go \textbf{Wrong}.

An attacker can cause traces of the ICS to diverge from the specified protocol. To account for this possibility, we implement dynamic protocol attestation. Dynamic protocol attestation checks that the ICS's trace conforms to the protocol. Dynamic protocol attestation prevents non-conforming RPCs from taking the system to a \textbf{Wrong} state.

\subsection{DSL Design}
We describe communication protocols using a DSL embedded in F*. Our methodology provides a general purpose protocol description language. However, the protocol-specific parameters listed in Table \ref{tab:parameter-description} must be provided by the engineer.

\begin{table}[h]
    \centering
    \begin{tabular}{|c|c|}
        \hline
        \textbf{Parameter} & \textbf{Description} \\ \hline
        $\mathbb{D}(V_m)$ & Values used by $State$ and $ExtCom$ \\ \hline
        $AP$ & Propositions that reflect the ICS's state \\ \hline
        $ExtCom$ & RPCs executed in the ICS \\ \hline
        Inputs  & Commands received by the HMI \\ \hline
    \end{tabular}
    \caption{Description of protocol-specific parameters.}
    \label{tab:parameter-description}
\end{table}

\subsubsection{Syntax}
Our DSL draws a distinction between RPCs and terms that describe protocol behavior. In our DSL, RPCs are called \emph{external commands}. Terms that describe protocol behavior are called \emph{internal commands}.

\[
\begin{aligned}
v_{0,j} &\in \mathbb{D}(V_0) \\
\vdots~ &\in \ldots \\
v_{m, j} &\in \mathbb{D}(V_m) \\
ap &\in AP \\
state &\in State \\
EC &\in ExtCom \\
IC &\in IntCom \\
\end{aligned}
\]

\[
\begin{array}{lll}
 V_0 &::= &v_{0,0}~ |~ \ldots~ |~ v_{0, n} \\
 ~\vdots &::= &\ldots \\
 V_m &::= &v_{m,0}~ |~ \ldots~ |~ v_{m, n} \\
 state & ::= &\textbf{ $\sigma$ $|$ Wrong} \\
 EC & ::= &E_1~ v_{m_1,n_1}~ \ldots~ v_{m_k,n_k}~ |~ \ldots~ |~ \\
    & &E_j~ v_{m_1,n_1}~ \ldots~ v_{m_k,n_k}  \\
 IC & ::= &\textbf{Skip}~ |~ \textbf{Seq}~ IC_1~ IC_2~ |~ \textbf{If}~ ap~ IC_1~ IC_2~ |~ EC
\end{array}
\]

$ExtCom$ is the set of all external commands exchanged in the system. Commands are represented as $E_j~ v_{m_1,n_1}~ \ldots~ v_{m_k,n_k}$. Each $v_{m_k,n_k}$ represents a value of the command's corresponding argument. A value is part of a domain $\mathbb{D}(V_m)$. Domains can be standardized (e.g., $\mathbb{N}$, bool) or user-defined. A trace is a sequence of external commands. External commands are the only mechanism to modify the ICS's state. Therefore, to assure the safety of the system is to enforce that only traces that satisfy safety guarantees are executed. In our methodology, this can be accomplished by defining a provably secure RPC protocol. Atomic propositions, $AP$, are used to reason about the ICS's internal state. The labeling function L: State $\to 2^{AP}$ maps a state to a set of atomic propositions. An atomic proposition $p$ evaluates to true in state $\sigma$ iff $p \in L(\sigma)$. 

States can be \textit{valid} or \textit{invalid}. A valid state is a tuple $\sigma = \langle s1, ~ s2, ~ \dots, ~ s_{k}, ~ input \rangle$; $\sigma$ contains $k$ internal variables that comprise the internal state. A valid state $\sigma$ also keeps track of SCADA and HMI input. Although not part of the syntax, inputs are also protocol-specific parameters. An invalid state is labeled as a \textbf{Wrong} state. Non-conformant RPCs take the semantics of the ICS to \textbf{Wrong}. Engineers define a protocol using internal commands. We emphasize the \textbf{If} internal command, as it represents the control flow for creating safe protocol specifications. \textbf{If} commands are conditioned by atomic propositions.

\subsubsection{Semantics}
\begin{figure*}[t!p]
\centering
\begin{minipage}{1\textwidth}
\scriptsize  
\begin{align*}
   \frac{
     \Phi(E_j v_{m_1,n_1} \ldots v_{m_k,n_k}, \sigma) \quad  \frac{
     \Phi(E_j v_{m_1,n_1} \ldots v_{m_k,n_k}, \sigma)  }{
     (E_j v_{m_1,n_1} \ldots v_{m_k,n_k}, \sigma) \Rightarrow_{EC} \sigma ' 
   } 
   }{
     (E_j v_{m_1,n_1} \ldots v_{m_k,n_k}, \sigma) \Rightarrow_{EC} \sigma ' 
   } \text{(ExtCom Valid)} \quad
   \frac{
   }{
     (E_j v_{m_1,n_1} \ldots v_{m_k,n_k},, \textbf{Wrong}) \Rightarrow_{EC} \textbf{Wrong}
   } \text{(Wrong)}  
\end{align*}

\begin{align*}
   &\frac{
     \neg  \Phi(E_j v_{m_1,n_1} \ldots v_{m_k,n_k}, \sigma)
   }{
     (E_j v_{m_1,n_1} \ldots v_{m_k,n_k}, \sigma) \Rightarrow_{EC} \textbf{Wrong}
   } \text{(ExtCom Invalid)} \quad
    \frac{
     ap \in L(s) \quad \quad (Ic1, s) \Rightarrow_{IC} s'
   }{
     (\textbf{If }ap ~IC_1 ~IC_2, s) \Rightarrow_{IC}  s'
   } \text{(If true)} \quad
      \frac{
   }{
     (\textbf{Skip}, s) \Rightarrow_{IC} s
   } \text{(Skip)} 
\end{align*}

 \begin{align*}
   &\frac{
     ap \notin L(s) \quad \quad (IC_2, s) \Rightarrow_{IC} s''
   }{
     (\textbf{If }ap ~IC_1~ IC_2, \sigma) \Rightarrow_{IC} s''
   } \text{(If false)} \quad
   \frac{
     (Ec1, s) \Rightarrow_{IC} \textbf{Wrong}
   }{
     (\textbf{Seq} ~IC_1 ~IC_2, s) \Rightarrow_{IC} \textbf{Wrong}
   } \text{(Seq Invalid)} \quad
   \frac{
     (IC_1, \sigma) \Rightarrow_{IC} \sigma' \lor (IC_2, \sigma) \Rightarrow_{IC} \sigma''
   }{
     (\textbf{Seq} ~IC_1 ~IC_2, \sigma) \Rightarrow_{IC} \sigma''
   } \text{(Seq Valid)} \quad
 \end{align*}

\caption{Semantics of our DSL.}
\label{fig:abstractsemantics}
\end{minipage}
\end{figure*}

The semantics in Figure \ref{fig:abstractsemantics} formalize the ICS's behavior. A command that takes a system to a valid state is a \textit{valid} command. Each external command has two semantic rules. The first rule states the effects of performing the operations denoted by a valid command. The second rule states which commands are \textit{invalid}. For example, the \textbf{Store} command is invalid when the control unit executes a \textbf{Store} command when the high bay warehouse is already full. Invalid commands lead the system to a \textbf{Wrong} state. The big-step semantics for external commands takes a state and a command and returns a new state.

\[
 \langle \text{ExtCom,~ State} \rangle \Rightarrow_{EC}   \text{State}  
\]

The semantics for internal commands also uses a big-step semantics. In the absence of recursion, all command-state pairs evaluate to a state without the need to consider intermediate steps.

\[
\langle \text{IntCom,~ State} \rangle \Rightarrow _{IC} \text{State} 
\]

In order to define a protocol specification for a given system, the user must add the system's external commands to the DSL. Furthermore, the user must define their underlying semantics. There are two functions that achieve these goals.

The semantic function \textbf{$\mathcal{E} : \text{ExtCom} \rightarrow \text{State} \rightarrow \text{State}$} implements $\Rightarrow_{EC}$ for all combinations of external commands and states. This function allows the user to define the semantics for all RPCs in the system. First, the semantics allows the system to define which $State \times ExtCom$ combinations are valid. Second, the semantics formalize how executing a given command changes the internal state. Third, the exhaustive property of F* functions forces users to consider all possible scenarios. However, some of the control flow may require procedures to check properties of the internal state.

The precondition function $\Phi : \text{ExtCom} \rightarrow \text{State} \rightarrow \text{bool}$ determines if a command is valid in a valid state. Commands that are not valid evaluate to \textbf{Wrong}. We note that $\Phi$ and atomic propositions are strictly orthogonal ideas. $\Phi$ is used to define the semantics of commands. Atomic propositions are used to define protocols. 

\begin{table*}[h]
\centering
    \begin{tabular}{|c|c|c|c|c|c|c|c|c|}
        \hline
            \multirow{2}{*}{\textbf{Framework}} & 
            \multirow{2}{*}{\textbf{Overhead}} & 
            \multicolumn{3}{|c|}{\textbf{Embedded Language Support}} & 
            \multirow{2}{*}{\textbf{Custom Networking?}} & 
            \multicolumn{2}{|c|}{\textbf{Security Practices}} \\
            \cline{3-5} \cline{7-8}  &  & \textbf{C?} & \textbf{C++?} & \textbf{Rust?} &  & \textbf{Code Scanning?} & \textbf{Fuzzing?}  \\
        \hline
        Cap'n Proto \cite{capnProto} & Low & \xmark & \checkmark & \checkmark & \xmark & \xmark & \checkmark \\
        COAP \cite{coap} & High & \checkmark & \checkmark & \checkmark & \xmark & \xmark & \xmark \\
        eRPC \cite{erpc} & Low & \checkmark & \xmark & \xmark  & \checkmark & \xmark & \xmark \\
        gRPC \cite{grpc} & Medium & \xmark & \checkmark & \checkmark  & \xmark & \xmark & \checkmark \\
        \hline
    \end{tabular}
    \caption{RPC frameworks we considered adopting in our prototype.}
    \label{tab:RPC-frameworks}
\end{table*}

\subsubsection{Safety Condition}\label{sec: safety property}
In an unrestricted environment, both valid and invalid sequences of RPCs can be executed. To generate only valid RPCs, we restrict accepted traces to a subset of all possible traces. Our DSL introduces syntax for restricting RPCs, given underlying semantics of the RPCs.  Using the DSL, users write a term (i.e., called $spec$) that specifies a communication protocol. F* can be used to prove different safety properties about the protocol. Given the attack model, the following safety condition must be proven:
\begin{center}
    
\fbox{
    \parbox{0.44 \textwidth}{
      \begin{center}
          \textbf{SAFETY CONDITION}
      \end{center}

      Applying the big-step semantics $\Rightarrow_{IC}$ to $spec$ in a state $ \ne \textbf{Wrong}$ always yields a state' $ \ne \textbf{Wrong}$.\\

      Formally:
      \[
      \forall \sigma \in \text{State}. \langle spec, \sigma \rangle \not \Rightarrow_{IC} \textbf{Wrong}
      \]
    }
}
\end{center}

If this condition is proven, then all sequences of commands generated by the $spec$ are safe command sequences. Thus, $spec$ generates a subset of all RPCs that cannot go \textbf{Wrong}. 

\subsection{Translation to RPC Framework}
We provide automation that translates terms from F* into an RPC framework. The translation occurs via procedures written using Meta-F* \cite{meta-fstar}. Meta-F* is a framework for  metaprogramming in F*. Metaprogramming involves writing procedures that write programs. Metaprogramming allows us to inspect terms in F* and emit new terms that generate code in an RPC framework. This allows us to use F* specifications to generate code for an RPC framework.

\subsubsection{RPC Frameworks}
Table \ref{tab:RPC-frameworks} shows the RPC frameworks we considered adopting in our prototype. The ``Custom Networking'' column indicates if the framework supports additional network technologies. There are a large number of network technologies found in embedded systems. For example, an embedded system might use CAN, UART, SPI, and ethernet to connect its subsystems. Ideally, an embedded framework allows for extending its network layer to support these protocols.

RPC frameworks introduce overhead. Overhead originates from serializing and deserializing data. We classified overhead as either \emph{Low}, \emph{Medium}, or \emph{High}. Cap'n Proto and eRPC are low-overhead RPC frameworks. They use binary representations of data that matches how a compiler arranges data structures. This reduces overhead while serializing and deserializing data. We consider COAP to have high overhead. While COAP itself is light-weight, it is not a true RPC framework. Creating an RPC framework in COAP introduces additional overhead beyond the other RPC frameworks.

We evaluated the security practices of candidate RPC frameworks with two criteria. First, \emph{does the project use any static analysis tools?} Static analysis tools like Coverity \cite{coverity-scan} can identify vulnerabilities during development \cite{coverity-sec}. Second, we considered \emph{does the project use fuzzing?} Fuzzing is a software testing strategy that uses random inputs to try to find unexpected outputs or invariant violations \cite{fuzzing}. Fuzzing has identified thousands of defects in software \cite{fuzzing-survey, fuzzing-roadmap}. The ``Security Practices'' columns in Table \ref{tab:RPC-frameworks} reports our findings.  gRPC performed static analysis using Coverity, but stopped in 2015. Both gRPC and Cap'n Proto use fuzzing. Fuzzing is effective, having found a critical security vulnerability in Cap'n Proto in 2022 \cite{capnproto-fuzzing}. 

We  chose to initially implement support for the Cap'n Proto RPC framework based on our evaluation. Cap'n Proto is lightweight, supports many embedded languages, and has good security practices. Although Cap'n Proto does not support C, C and C++ have good interoperability.

In the future, we plan to support additional RPC frameworks besides Cap'n Proto. Cap'n Proto lacks the ability to use common embedded networking technologies. We will eventually fill this gap using eRPC. In the long term, we plan to create our own RPC framework. We will use formal methods to prove its security. 

\subsection{Dynamic Protocol Attestation}

\begin{figure}[H]
    \centering
    \includegraphics[width=1\linewidth]{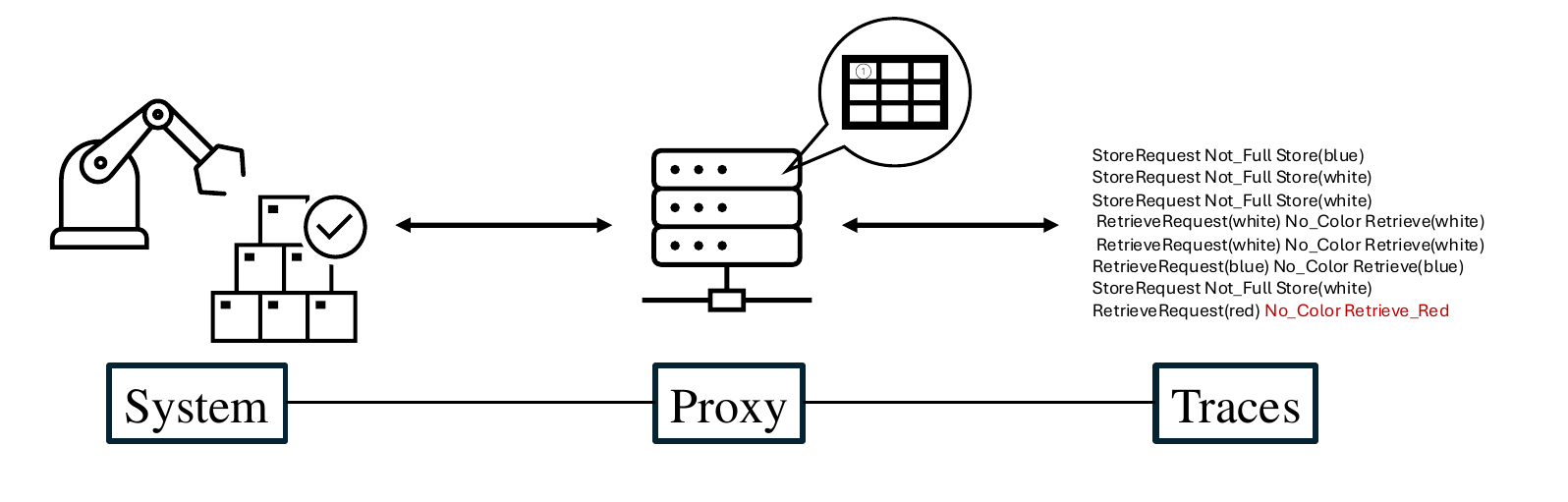}
    \caption{Dynamic protocol attestation.}
    \label{fig:dynamic}
\end{figure}

We make no assumptions about the correctness of components implementing the RPC protocol beyond the low-level device drivers. Component defects may produce invalid traces. Thus, we introduce dynamic protocol attestation to ensure the protocol is followed at runtime. Consequently, it provides assurance that the system does not reach an unsafe state. To prevent illegal runtime traces from causing the ICS to go wrong, we must decide if the current trace is generated by $spec.$ A decider procedure acts as a recognizer for the safe RPC language. The decider takes the ongoing system trace, $spec$, and current state as an input. As output, the decider returns true if $spec$ generates the trace and false otherwise. This procedure is implemented as a proxy. The proxy monitors ICS traces at runtime, as shown in Figure \ref{fig:dynamic}. For  performance reasons, we implemented the proxy by hand. But, it can be automatically generated. This automation is left for future work.

\section{Applying Methodology to Fischertechnik}

\begin{figure}[H]
    \centering
    \includegraphics[width=1.0\linewidth]{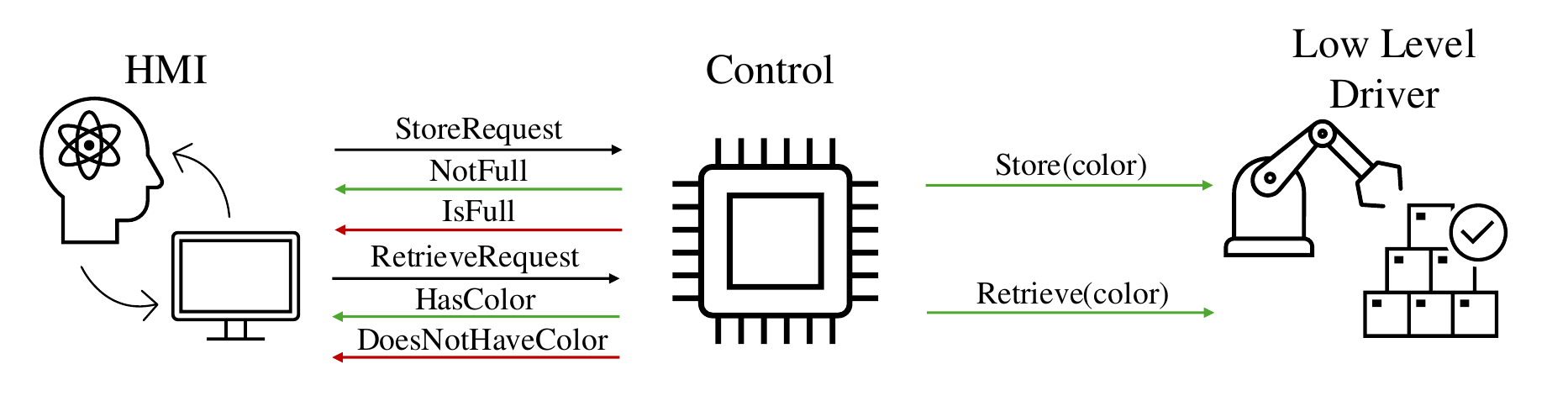}
    \caption{The warehouse's communication protocol.}
    \label{fig:ft-example}
\end{figure}

To demonstrate our metholodology, we use the Fischertechnik factory as shown in Figure \ref{fig:ft-example}.  A customer can request an item of a certain color via a \textbf{RetrieveRequest} $color$ RPC. Likewise, a supplier can request to store an item using the \textbf{StoreRequest} $color$ RPC. The HBW controller receives and processes these RPCs. The effect of RPCs is contingent on the current internal state of the system. The controller can either return a rejection or an acknowledgment to the sender. The \textbf{RetrieveRequest} $color$ RPC triggers either a \textbf{HasColor} or a \textbf{DoesNotHaveColor} response. The \textbf{StoreRequest} $color$ RPC can trigger a \textbf{IsFull} or \textbf{NotFull} response. The controller then forwards the accepted requests to the low-level hardware drivers. After processing an RPC, the controller updates the current dynamic state of the the bays. 

The controller makes all policy decisions. The low-level drivers faithfully execute all RPCs they receive from the controller. The safety of this communication is critically important. An attacker is able to execute arbitrary RPCs. Thus, an attack stemming from the controller could trigger unsafe driver commands. These unsafe commands pose a physical threat to workers, and even the ICS itself. Since the protocol describes all safe executions, nonconforming traces can cause unsafe behavior. Non-conformant traces that lead to unsafe behavior are \emph{invalid traces}. A invalid trace will break at least one logical assertions defined in \S \ref{sec:safety-properties}. This can lead to outcomes like collisions when attempting to store an item in an occupied bay. 


\subsection{Syntax and Semantics}\label{sec: FT proofs}

\begin{figure*}[t!p]
\centering
\begin{minipage}{1\textwidth}
\scriptsize 
\begin{align*}
  &\frac{
    \text{find color}(s,1,\text{empty}) = \text{None}
  }{
    (\textbf{IsFull},s) \Rightarrow_{EC} s
  } \text{(IsFull Valid)} \quad
  \frac{
    \text{find color}(s,1,\text{empty}) = \text{Some }
  }{
    (\textbf{IsFull},s) \Rightarrow_{EC} \textbf{Wrong}
  } \text{(IsFull Invalid)} \quad
  \frac{
    \text{find color}(s,1,\text{empty}) = \text{Some }
  }{
    (\textbf{NotFull},s) \Rightarrow_{EC} s
  } \text{(NotFull Valid)} 
\end{align*}

\begin{align*} 
  &\frac{
    \text{find color}(s,1,\text{empty}) = \text{None}
  }{
    (\textbf{NotFull},s) \Rightarrow_{EC} \textbf{Wrong}
  } \text{(NotFull Invalid)} \quad
  \frac{
    \text{find color}(\sigma, 1, \text{empty}) = \text{None}
  }{
    (\textbf{Store}~color, \sigma) \Rightarrow_{EC} \textbf{Wrong}
  } \text{(Store color) Invalid)} \quad
  \frac{
    \text{find color}(\sigma, 1, \text{color}) = \text{None}
  }{
    (\textbf{Retrieve}~color, \sigma) \Rightarrow_{EC} \textbf{Wrong}
  } \text{(Retrieve Color Invalid)}
\end{align*}

\begin{align*}
  &\frac{
    \sigma.\text{inp} = \textbf{RetrieveRequest } color \quad \quad \text{find color}(\sigma, 1, c) = \text{None}
  }{
    (\textbf{DoesNotHaveColor}, \sigma) \Rightarrow_{EC} \sigma
  } \text{(DNHC Valid)} \quad
\frac{
    \sigma.\text{inp} = \textbf{RetrieveRequest } color \quad \quad \text{find color}(\sigma, 1, c) = \text{Some }
  }{
    (\textbf{HasColor}, \sigma) \Rightarrow_{EC} \sigma
  } \text{(HasColor Valid)} 
\end{align*}

\begin{align*}
   &\frac{
    \sigma.\text{inp} = \textbf{RetrieveRequest } color \quad \quad \text{find color}(\sigma, 1, c) = \text{Some }
  }{
    (\textbf{DoesNotHaveColor}, \sigma) \Rightarrow_{EC} \textbf{Wrong}
  } \text{(DNHC Invalid)} \quad
  \frac{
    \sigma.\text{inp} = \textbf{RetrieveRequest } color \quad \quad \text{find color}(\sigma, 1, c) = \text{None}
  }{
    (\textbf{HasColor}, \sigma) \Rightarrow_{EC} \textbf{Wrong}
  } \text{(HasColor Invalid)} 
\end{align*}

\begin{align*}    
  &\frac{
    \text{find color}(\sigma, 1, \text{empty}) = \text{Some }
  }{
    (\textbf{Store}~color, \sigma) \Rightarrow_{EC} \sigma [\text{color} \mapsto \text{bf}(\sigma\text{.bays}, n)] 
  } \text{(Store color Valid)} \quad 
  \frac{
    \text{find color}(\sigma, 1, \text{color}) = \text{Some n}
  }{
    (\textbf{Retrieve}~color, \sigma) \Rightarrow_{EC} \sigma [\text{empty} \mapsto \text{bf}(\sigma\text{.bays}, n)]  
  } \text{(Retrieve color Valid)} \quad
\end{align*}

\caption{Semantics of our command language.}
\label{fig:semantics}
\end{minipage}
\end{figure*}

The protocol-specific syntax is:

\[
\begin{array}{lll}
 color & ::= &\textbf{empty $|$ red $|$ white $|$ blue} \\
 input & ::= &\textbf{RetrieveRequest $color$ $|$ StoreRequest $color$}  \\
 EC & ::= &\textbf{ IsFull $|$ NotFull $|$ HasColor $|$ Store}~ color \\
    & &~ \textbf{$|$ Retrieve}~ color | \textbf{DoesNotHaveColor}  \\
 ap & ::= &\textbf{APBaysFull} ~ | ~ \textbf{APHasRetrieveRequest}  ~  \\
&& ~ | ~ \textbf{APContainsWhite}  ~ | ~ \textbf{APContainsBlue} \\
  && ~ | ~ \textbf{APHasStoreRequest}
  ~ | ~ \textbf{APRequestRed}  \\
   &&~ | ~ \textbf{APContainsRed} ~ | ~ \textbf{APRequestBlue} \\
 &&~ | ~\textbf{APRequestWhite}
\end{array}
\]

Some external commands take $color$ as arguments.  We represent a valid state as a pair $\sigma = \langle$ bays, input $\rangle$; $\sigma$ keeps track of the current state of the nine bays and the current user input. We model the bays as a record of nine slots:

\begin{lstlisting}[style=fstar,frame=none,numbers=none]
type bays = { bay1 : color; $\dots$ ; bay9 : color; }
\end{lstlisting}

To access each bay, we define a function:

\begin{lstlisting}[style=fstar,frame=none,numbers=none]
    bf : bays $\rightarrow$ $\mathbb{N}$ $\rightarrow$ color
\end{lstlisting}
that takes the bays and a number $n$ between 1 and 9 as arguments, and returns the current state of bay $n$.

$EC$ is the set of valid commands that can be sent in the network. Figure \ref{fig:ft-example} demonstrates the intended communication protocol.  The example in appendix \S \ref{app: valid traces} shows how a trace is broken down into small sequences of commands that can be interpreted by the decider. Each part is evaluated by the dynamic attestation checker against the current state. The example contains only valid commands that take the HBW from a valid state to another valid state according to the HMI input. On the other hand, \S \ref{app: Invalid traces} illustrates a few patterns of invalid traces. These traces contain at least one invalid command that takes the ICS state to \textbf{Wrong}. The invalid command either violates the HBW's safety properties listed in \S \ref{sec:safety-properties} or does not correspond to the intended action given by the HMI input. 

Using our DSL, we write a term called $spec$ (which is shown in Figure \ref{fig:spec}) that specifies the RPC protocol for the high bay warehouse. 

\begin{figure}
\begin{lstlisting}[gobble=3, style=fstar,frame=none]
   $spec$ =
   If APHasStoreRequest 
    (If APBaysFull (IsFull)
      (Seq 
        (NotFull)
        (If APRequestRed (Store Red)
          (If APRequestWhite (Store White)
            (If APRequestBlue (Store Blue)
              Skip)))))
    (If APHasRetrieveRequest
      (If APRequestRed
        (If APContainsRed 
          (Seq (HasColor) (Retrieve Red)) (NoColor))
        (If APRequestWhite
          (If APContainsWhite 
            (Seq (HasColor) (Retrieve White)) 
            (NoColor))
          (If APRequestBlue
            (If APContainsBlue 
              (Seq (HasColor) (Retrieve Blue))
              (NoColor))
            Skip)))
       Skip)
\end{lstlisting}
\caption{Specification of the HBW's protocol.}
\label{fig:spec}
\end{figure}

Next, we must prove that following the protocol described by $spec$ ensures that the safety condition holds. For that purpose, we define the semantic function for the RPC language. 

To assist with the controller logic, we define the function:
\begin{lstlisting}[style=fstar,frame=none,numbers=none]
    find_color : color $\rightarrow$ $\mathbb{N}$ $\rightarrow$ bays $\rightarrow$ option $\mathbb{N}$
\end{lstlisting}
which acts as the $\Phi$ function. \lstinline{find_color empty b} finds the bay that will be used during a \textbf{Store} RPC. Similarly, \lstinline{find_color c b} finds a item colored \lstinline{c}. The function returns an \lstinline{option $\mathbb{N}$ }. Values of type \lstinline{option a} are either \lstinline{Some a} or \lstinline{None}. The natural number inside the option cannot be used without accounting for the possibility that the value is missing. With the help of the precondition function, we define the semantic function in \S \ref{app:sem-fun}. To define a semantic function, the user must match each external command to its valid and invalid semantic rules. As an example of conformant RPC behavior, an \textbf{IsFull} response can be sent when the bays are full. On the other hand, an \textbf{IsFull} response sent when one more bay is available is an invalid behavior. The semantic function formalizes these semantic rules. Figure \ref{fig:semantics} shows the semantics rules of the HBW commands. The semantic rules will serve as the basis for the safety condition proof.

The proof in \S \ref{app:safety-proof} verifies our safety condition. The proof shows that, for any valid state $\sigma$, $\langle spec, \sigma \rangle \not \Rightarrow_{IC} \textbf{Wrong}$. The proof verifies that every combination of atomic propositions allowed by $spec$ in a valid state cannot compute a \textbf{Wrong} state. A \lstinline{match} construct in F* relates atomic propositions to the internal state. As an example, if the labeling function maps a state to APHasStoreRequest and APBaysFull, then the HBW must be full. Otherwise, APBaysFull means we sent an invalid command to the HMI.  An \lstinline{assert} construct in F* checks that to be true. Unlike other ITPs, F* can delegate the verification of boolean equations to SMT solvers. F* uses an SMT solver backend to alleviate some of the proof burden. Instead of writing hints to the proof checker, engineers write assertions they want to verify. These assertions are encoded as SMT formulae that can be checked by an SMT solver. The proof uses the flexibility of this approach. Assertions are checked at compile-time, before the program is executed. The proof is complete once all atomic propositions are checked to reach valid states when adhering to the $spec$ protocol.

Consider a situation where there is a request to retrieve a blue item. Then, a system that follows $spec$ denies the request by sending a \textbf{NoColor} RPC. The safety proof guarantees that there is no blue item in the high bay warehouse. The proof uses the the labeling function $L$ to map the current state to the atomic propositions that are true at that state. Then it uses an assert statement to ask the Z3 solver to verify APContainsBlue $\in\ L(\sigma)$. The SMT backend makes this proof style more ergonomic then solely relying on type-checker. 

We provide the implementation of a decider that checks traces are generated by a protocol in F*, as shown in \S \ref{app:decider}. The decider uses the \lstinline{eval_until_next_com} function to break down the trace into smaller command sequences. This function is responsible for breaking down traces into checkable RPC sequences.

Given the proven safety condition, $spec$ cannot lead to a \textbf{Wrong} state. The decider then makes is possible to decide if a trace can be generated by $spec$. The decider will be used to implement the dynamic protocol attestation checker discussed in \ref{sec: dyn attest}.

\subsection{RPC Generation for the Fischertechnik}
\begin{figure*}
    \centering
    \begin{subfigure}[t]{0.46 \textwidth}
        \begin{lstlisting}[style=fstar,frame=none]
type store_response = | NotFull | IsFull

type retrieve_response = 
  | HasColor | DoesNotHaveColor

type color = | Red | White | Blue

type hbw_rpc = {
    store_request : color $\rightarrow$ store_response;
    retrieve_request : color $\rightarrow$
        retrieve_response;  
    store : color $\rightarrow$ unit;
    retrieve : color $\rightarrow$ unit;
}

let enc : encoder CapnProto hbw_rpc =
    ~\label{listing:gen_rpcs}~ _ by (gen_rpcs (`CapnProto) (`hbw_rpc))

dump enc
\end{lstlisting}
        \caption{The RPC specification in F*.}
        \label{listing:rpc-definition}
    \end{subfigure}
   \begin{subfigure}[t]{0.46 \textwidth}
        \begin{lstlisting}[style=fstar,frame=none]
interface HbwRpc {
  storeRequest @0(_0:Color) -> StoreResponse;
  retrieveRequest 
      @1(_0:Color) -> RetrieveResponse;
  store @2(_0:Color);
  retrieve @3(_0:Color);
}
struct RetrieveResponse {
  enum V {hasColor @ 0; doesNotHaveColor @ 1;}
  v @ 0 : V;
}
struct StoreResponse {
  enum V {notFull @ 0; isFull @ 1;}
  v @ 0 : V;
}
struct Color {
  enum V {red @ 0; white @ 1; blue @ 2;}
  v @ 0 : V;
}
\end{lstlisting}
        \caption{Cap'n Proto inputs emitted from Figure \ref{listing:rpc-definition}.}
        \label{listing:capn-proto-code}
    \end{subfigure}
    \caption{Overview of the Fischertechnik's HBW RPC system.}
\end{figure*}

Figure \ref{listing:rpc-definition}  shows F* code that describes the RPCs comprising our protocol. Critically, line \ref{listing:gen_rpcs} shows how to use our metaprogramming system. The \lstinline{by} syntax invokes F*'s tactic system. F*'s tactic system is an extensible mechanism to automate transforming terms. Individual term transformers are called \emph{tactics}. \lstinline{gen_rpcs} is a tactic that transforms an RPC framework and an F* type into an RPC encoder. RPC encoders contain code to emit inputs to an RPC framework. Emission occurs by calling the \lstinline{dump} function. Figure \ref{listing:capn-proto-code} shows the inputs that are automatically generated from the code in Figure \ref{listing:rpc-definition}.

The tactic system manipulates F*'s abstract syntax tree (AST) using \lstinline{gen_rpcs}. \lstinline{gen_rpcs} traverses the AST of the type supplied as its second argument. Several primitive types including machine integers and strings have simple, direct encodings. \lstinline{gen_rpcs} supports composite types like records by recursively applying itself to struct members. Advanced features in F*'s type system like sum types and refinement types are unsupported. 

The RPC framework's existing code generation tool transforms the framework inputs into code. We trust the RPC framework to emit correct code. The generated code contains logic to handle errors ranging from RPC cancellation to network errors. Since common errors are already handled, engineers only need to implement message semantics. 

\subsection{Dynamic Protocol Attestation}
\label{sec: dyn attest}

 We implement dynamic protocol attestation as a proxy in C++ using Cap'n Proto. The proxy dynamically feeds the current trace to the decider procedure in F*. Since ICSs never terminate, our approach ensures that we always analyze finite traces. That is done through chopping the trace in smaller parts while keeping track of the state. Our approach was successful in finding non-comformant RPCs.

\subsubsection{Performance Evaluation}
To understand the effects of dynamic protocol attestation on RPC performance, we conducted a performance evaluation. To conduct our evaluation, we created a software implementation of the Fischertechnik HBW system. Our implementation has four components: a controller, low-level device drivers, a client, and the dynamic attestation system.  The controller exposes a store/retrieve interface to the client. The low-level device drivers executes physical actions. The client randomly sends the controller messages that conform to the controller's protocol. We implemented each component in C++. We implemented RPCs using Cap'n Proto with the inputs our automation generated. 

We considered two performance criteria in our evaluation. First, \emph{what is the effect on latency?} Second, \emph{what is the effect on message throughput?} We examined two measures of throughput: the number of messages sent, and the number of bytes sent. We evaluated two configurations: one without dynamic protocol attestation enabled, and one with it enabled. We evaluated each configuration for 30 seconds. Table \ref{tab:perf_results} shows our numeric results. 

\begin{table}[h]
    \centering
    \begin{tabular}{|c|c|c|c|}
          \hline
         \multirow{2}{*}{\textbf{Configuration}} & \multirow{2}{*}{\textbf{\textbf{Avg. Latency (ms)}}}  & \multicolumn{2}{|c|}{\textbf{Throughput}} \\
         \cline{3-4} &  & \textbf{Messages} & \textbf{KB} \\
         \hline
         Without Attestation & $0.1452 \pm 0.0007$ & 191,250 & 3,060 \\
         With Attestation & $0.1903 \pm 0.0007$ & 145,054 & 2,321 \\
         \hline
    \end{tabular}
    \caption{Numeric results from our
    evaluation.}
    \label{tab:perf_results}
\end{table}

There is moderate overhead imposed by protocol attestation. Protocol monitoring increases latency by $\approx 31\%$. Similarly, protocol monitoring decreases throughput by $\approx 32\%$. Some applications may tolerate protocol monitoring's performance overhead. Performance overhead can be reduced in future work.

\begin{figure}
    \centering
    \includegraphics[scale=.75]{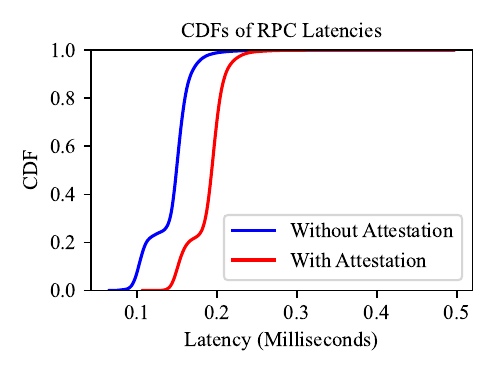}
    \caption{CDFs of RPC latency, without and with dynamic protocol attestation.}
    \label{fig:monitoring_cdf}
\end{figure}

To better understand dynamic protocol attestation's impact on latency, we examined the cumulative density functions (CDFs) of our two evaluation configurations. These CDFs are shown in Figure \ref{fig:monitoring_cdf}. We observe that protocol monitoring preserves the shape of the original CDF. In particular, $99^{\text{th}}$ percentile latency also increases by $\approx 30\%$. Despite this increase, it is promising that protocol monitoring's worst-case performance appears tied to its average-case performance.
\section{Limitations}
In this section we will briefly discuss some of the known limitations of our work and possible mitigations for them. The protocols defined with our methodology provide formal guarantees of safety. However, these guarantees exist within a framework of assumptions.

\subsection{Fail-Safes Are Not Always Safe}
The methodology guarantees that non-conformant RPCs will not execute dangerous commands. Instead, the ICS executes a fail-safe command whenever an RPC would violate the protocol. 
Often these fail-safe commands are well-known and can be safely executed from any given state the ICS may be in. However, there may be ICSs for which such fail-safes do not exist. Moreover fail-safes may rely on the integrity of the devices that trigger them. Possibly, an alternative backup mechanism could be called when a fail-safe needs to be executed. One such mechanism is to alert a human operator to manually bring the ICS to a known safe state. 

Additionally, we assume doing nothing is always safe. This assumption has the unintended consequence of requiring our controller to be sufficiently abstracted away from the drivers. Then, issuing no commands or issuing a fail-safe is always a safe operation. Some ICS subsystems rapidly issue commands turning another subsystem on and off at a precise interval (e.g., a pulse-width modulated controller). 
In the event the controller must execute a fail-safe command, it may not be safe to cease issuing other commands. The controller could have been actively preventing another component from causing damage. Our protocol makes no guarantees about the timeliness of the messages between devices. Thus, these time-dependent events should be pushed into the devices' drivers.

\subsection{No Liveliness Guarantees; Denial of Service} 
We assume no guarantees about time within our system nor any liveliness guarantees. 
A compromised subsystem could continuously execute non-conformant RPCs. Consequently, the only executed commands are fail-safes. Alternatively, the compromised device could refuse to communicate at all, denying service to all devices requiring a response from it. A possible defense to this is to have an internal timeout process within the protocol. The timeout process would issue a competing command to a controller, instructing the controller to issue a fail-safe within a reasonable amount of time. Such instruction can be implemented by refining the communication semantics.

\subsection{Discretization}
Our RPC functions allow any type within the function parameters. However, our type system can only reason about countable states and the protocol must be implemented on realizable hardware. 
This prevents us from reasoning about ICSs with possibly uncountable states or RPCs where infinite precision is required. However, we argue these limitations exist in the current ICS landscape. Furthermore, our solution drastically improves the assurances ICS operators can have on their system. 

\section{Future Work \& Conclusion}

\subsection{Future Work}

Our initial results demonstrate the viability of our approach. We envision four extensions to our work. First, the current DSL to define $spec$ requires protocol-specific parameters. Our ultimate goal is to create a general DSL that can describe all protocols. Second, the dynamic checker's overhead might prevent its adoption in certain scenarios. Third, most of the framework in this paper was implemented by hand. Additional automation is needed to create the dynamic checker. We also desire proof automation. Finally, we desire supported for additional RPC frameworks.



\subsubsection{Generalizing the DSL}

To create DSLs for defining arbitrary protocols, we need a core language. Session types \cite{honda1993} are a typed foundation for the design of message-passing systems. Multiparty session types (MPST)\cite{hondaMultiPartyAsync} are able to define properties in systems with multiple components. One of these properties is the ordering of messages. 
A relevant example is the application of MPST to specify the protocol for the Ocean Observatories Initiative (OOI) \cite{spy,Bocchi2017,Scribble}. This research inspired the development multiparty session types with payload refinements \cite{Zhou2020,vassor_RMPST,SessionRefFsharp}. Refinement multiparty session types (RMPSTs) are able to also reason about the contents of messages. A possible extension of our work is to embed RMPSTs in F*. This embedding allows users to write typed protocol specifications using RMPSTs. This can be seen a direct extension of \cite{Zhou2020}. Once having embedded RMPSTs in F*, we plan to integrate with additional type theories including theories of unit types and security type systems (e.g., \cite{sa4u,cocoon}).



\subsubsection{Reducing Dynamic Attestation's Overhead}
Our implementation of dynamic protocol attestation is naive. A single, global proxy adds performance overhead because it requires an extra round of serializing and deserializing RPCs. This extra round could be prevented if the proxy deserialized RPCs into shared memory. Deserializing RPCs into shared memory requires a distributed attestation system. We leave this as future work.


\subsubsection{Extending Automation}
To provide better automation for our approach, we could also benefit from session types. The use of a DSL that implements RMPSTs would mechanize the process in many ways. First, standard semantics are used for proofs. Second, F*'s type checker can be used to verify protocol implementations. Third, a few core proofs, such as absence of deadlocks, can be proven for all protocols implemented using our methodology. These proofs would be protocol-irrelevant. Users would get these benefits just by using the DSL. 

We have shown F*'s use of SMT solvers automates some of the proof process. However, considerable knowledge about F* is needed to prove non-trivial properties of complex system. This is a drawback that precludes adoption of formal methods. We can help reduce proof burden by (1) integrating with model checkers when possible, (e.g., \cite{bfi, avis}) and (2) providing new tactics and SMT patterns to help further automate proofs.


\subsubsection{Additional RPC Framework Support}
This approach currently only supports the Cap'n Proto RPC framework.  We would like to expand our extraction of RPCs to handle additional RPC frameworks. 
 
\subsection{Conclusion}

In this paper, we have presented a methodology that enables engineers to design communication protocols in F*. Communication protocols are provably safe using F*'s theorem proving capabilities. We have also shown how to translate the F* specification into RPCs. By extracting RPCs from the F* specification, we provide a single source of truth governing the ICS's RPCs. We implemented a dynamic attestation system that ensures runtime traces conform to the safe protocol. Critically, our approach limits the impact of compromised components without requiring heavy-weight verification. While these tradeoffs are not insignificant, many ICS could benefit from our proposal, especially when the cost of security incidents is high.


\bibliography{max_bib}

\begin{thebibliography}{10}

\bibitem{Omar2021}
Omar Al-Bataineh, Daniel Jun~Xian Ng, and Arvind Easwaran.
\newblock Monitoring cumulative cost properties.
\newblock In {\em Proceedings - 2021 IEEE/ACM 9th International Conference on Formal Methods in Software Engineering, FormaliSE 2021}, pages 19--30. Institute of Electrical and Electronics Engineers Inc., 5 2021.

\bibitem{coverity-sec}
Dejan Baca.
\newblock Identifying security relevant warnings from static code analysis tools through code tainting.
\newblock In {\em 2010 International Conference on Availability, Reliability and Security}, pages 386--390. IEEE, 2010.

\bibitem{model2}
Sebastian Biallas.
\newblock Verification of programmable logic controller code using model checking and static analysis.
\newblock Technical report, RWTH Aachen University, 2016.

\bibitem{Bocchi2017}
Laura Bocchi, Tzu~Chun Chen, Romain Demangeon, Kohei Honda, and Nobuko Yoshida.
\newblock Monitoring networks through multiparty session types.
\newblock {\em Theoretical Computer Science}, 669:33--58, 3 2017.

\bibitem{capnproto-fuzzing}
{Cap'n Proto}.
\newblock Cve-2022-46149 security advisory, 2022.
\newblock Accessed: 2024-09-03.

\bibitem{capnProto}
{Cap'n Proto Authors}.
\newblock Cap'n proto: Insanely fast data interchange format and rpc system.
\newblock \url{https://capnproto.org}, 2024.
\newblock Accessed: 2024-06-06.

\bibitem{Cattel1995}
Thierry Cattel.
\newblock Process control design using spin, 1995.

\bibitem{Cattel1996}
Thierry Cattel.
\newblock Using concurrency and formal methods for the design of safe process control, 1996.

\bibitem{coverity-scan}
{Coverity Scan}.
\newblock Coverity scan: Static analysis for open source projects, 2024.
\newblock Accessed: 2024-09-03.

\bibitem{z3}
Leonardo De~Moura and Nikolaj Bj{\o}rner.
\newblock Z3: An efficient smt solver.
\newblock In {\em International conference on Tools and Algorithms for the Construction and Analysis of Systems}, pages 337--340. Springer, 2008.

\bibitem{erpc}
{eRPC Project}.
\newblock erpc: Embedded remote procedure call, 2024.
\newblock Accessed: 2024-09-03.

\bibitem{applyingModelChecking}
Borja Fern{\'a}ndez~Adiego, Ignacio~D Lopez-Miguel, Frederic Havart, Enrique Blanco~Vi{\~n}uela, Tomasz Ladzinski, and Jean-Charles Tournier.
\newblock Applying model checking to highly-configurable safety critical software: the sps-pps plc program.
\newblock {\em JACoW}, pages 759--763, 2022.

\bibitem{FT}
Fischertechnik.
\newblock Fischertechnik.
\newblock \url{https://www.fischertechnik.de/en/}, 2024.
\newblock Accessed: 2024-06-06.

\bibitem{fischertechnikHBW}
Fischertechnik.
\newblock Fischertechnik automated high bay warehouse 24v, 2024.
\newblock Accessed: 2024-10-08.

\bibitem{Gourcuff2008}
Vincent Gourcuff, Jean~Marc Faure, and Olivier~De Smet.
\newblock Improving large-sized plc programs verification using abstractions.
\newblock In {\em IFAC Proceedings Volumes (IFAC-PapersOnline)}, volume~17, 2008.

\bibitem{grpc}
{gRPC Authors}.
\newblock grpc: A high performance, open source universal rpc framework.
\newblock \url{https://grpc.io}, 2024.
\newblock Accessed: 2024-06-06.

\bibitem{Haque2018}
Mohammad~Shihabul Haque, Daniel Jun~Xian Ng, Arvind Easwaran, and Karthikeyan Thangamariappan.
\newblock Contract-based hierarchical resilience management for cyber-physical systems.
\newblock {\em Computer}, 51:56--65, 11 2018.

\bibitem{honda1993}
Kohei Honda.
\newblock Types for dyadic interaction*.
\newblock 1993.
\newblock Accessed: 2024-09-03.

\bibitem{Scribble}
Kohei Honda, Aybek Mukhamedov, Gary Brown, Tzu-Chun Chen, and Nobuko Yoshida.
\newblock Scribbling interactions with a formal foundation.

\bibitem{hondaMultiPartyAsync}
Kohei Honda, Nobuko Yoshida, and Marco Carbone.
\newblock Multiparty asynchronous session types.
\newblock In {\em Proceedings of the 35th annual ACM SIGPLAN-SIGACT symposium on Principles of programming languages}, pages 273--284, 2008.

\bibitem{curryHoward}
William~A Howard et~al.
\newblock The formulae-as-types notion of construction.
\newblock {\em To HB Curry: essays on combinatory logic, lambda calculus and formalism}, 44:479--490, 1980.

\bibitem{ibm_industry_4_0}
{IBM}.
\newblock Industry 4.0, n.d.
\newblock Accessed: 2024-08-28.

\bibitem{bfi}
Saurabh Jha, Subho Banerjee, Timothy Tsai, Siva~KS Hari, Michael~B Sullivan, Zbigniew~T Kalbarczyk, Stephen~W Keckler, and Ravishankar~K Iyer.
\newblock Ml-based fault injection for autonomous vehicles: A case for bayesian fault injection.
\newblock In {\em 2019 49th annual IEEE/IFIP international conference on dependable systems and networks (DSN)}, pages 112--124. IEEE, 2019.

\bibitem{sel4Paper}
Gerwin Klein, Kevin Elphinstone, Gernot Heiser, June Andronick, David Cock, Philip Derrin, Dhammika Elkaduwe, Kai Engelhardt, Rafal Kolanski, Michael Norrish, et~al.
\newblock sel4: Formal verification of an os kernel.
\newblock In {\em Proceedings of the ACM SIGOPS 22nd symposium on Operating systems principles}, pages 207--220, 2009.

\bibitem{cocoon}
Ada Lamba, Max Taylor, Vincent Beardsley, Jacob Bambeck, Michael~D Bond, and Zhiqiang Lin.
\newblock Cocoon: Static information flow control in rust.
\newblock {\em Proceedings of the ACM on Programming Languages}, 8(OOPSLA1):166--193, 2024.

\bibitem{fuzzing-survey}
Jun Li, Bodong Zhao, and Chao Zhang.
\newblock Fuzzing: a survey.
\newblock {\em Cybersecurity}, 1:1--13, 2018.

\bibitem{plcVerif}
Ignacio~D Lopez-Miguel, Jean-Charles Tournier, and Borja~Fernandez Adiego.
\newblock Plcverif: status of a formal verification tool for programmable logic controller.
\newblock {\em arXiv preprint arXiv:2203.17253}, 2022.

\bibitem{meta-fstar}
Guido Mart{\'\i}nez, Danel Ahman, Victor Dumitrescu, Nick Giannarakis, Chris Hawblitzel, C{\u{a}}t{\u{a}}lin Hri{\c{t}}cu, Monal Narasimhamurthy, Zoe Paraskevopoulou, Cl{\'e}ment Pit-Claudel, Jonathan Protzenko, et~al.
\newblock Meta-f: Proof automation with smt, tactics, and metaprograms.
\newblock In {\em European Symposium on Programming}, pages 30--59. Springer International Publishing Cham, 2019.

\bibitem{steelMill}
Dover Microsystems.
\newblock German steel mill cyberattack, 2024.
\newblock Accessed: 2024-09-10.

\bibitem{fuzzing}
Barton~P Miller, Lars Fredriksen, and Bryan So.
\newblock An empirical study of the reliability of unix utilities.
\newblock {\em Communications of the ACM}, 33(12):32--44, 1990.

\bibitem{hodaOntology}
Ramesh Neupane and Hoda Mehrpouyan.
\newblock An ontology-based framework for formal verification of safety and security properties of control logics.
\newblock In {\em 2022 14th International Conference on Electronics, Computers and Artificial Intelligence (ECAI)}, pages 1--8. IEEE, 2022.

\bibitem{SessionRefFsharp}
Rumyana Neykova, Raymond Hu, Nobuko Yoshida, and Fahd Abdeljallal.
\newblock A session type provider: compile-time api generation of distributed protocols with refinements in f\#.
\newblock In {\em Proceedings of the 27th International Conference on Compiler Construction}, CC '18, page 128–138, New York, NY, USA, 2018. Association for Computing Machinery.

\bibitem{spy}
Rumyana Neykova, Nobuko Yoshida, and Raymond Hu.
\newblock Spy: Local verification of global protocols, 2013.

\bibitem{cic}
Christine Paulin-Mohring.
\newblock Introduction to the calculus of inductive constructions.
\newblock {\em All about Proofs, Proofs for All}, 55, 2015.

\bibitem{modelCheckingPlcFbds}
Olivera Pavlovic and Hans-Dieter Ehrich.
\newblock Model checking plc software written in function block diagram.
\newblock In {\em 2010 Third International Conference on Software Testing, Verification and Validation}, pages 439--448. IEEE, 2010.

\bibitem{stuxnet}
Michael Riley.
\newblock The real story of stuxnet.
\newblock {\em IEEE Spectrum}, 2014.
\newblock Accessed: 2024-09-10.

\bibitem{coap}
Z.~Shelby, C.~Bormann, L.~Frank, S.~Kasap, P.~Kyzivat, and J.~M.~D. Morris.
\newblock The constrained application protocol (coap), 2014.
\newblock RFC 7252, June 2014. Accessed: 2024-09-03.

\bibitem{fstar}
Nikhil Swamy, C{\u{a}}t{\u{a}}lin Hri{\c{t}}cu, Chantal Keller, Aseem Rastogi, Antoine Delignat-Lavaud, Simon Forest, Karthikeyan Bhargavan, C{\'e}dric Fournet, Pierre-Yves Strub, Markulf Kohlweiss, et~al.
\newblock Dependent types and multi-monadic effects in f.
\newblock In {\em Proceedings of the 43rd annual ACM SIGPLAN-SIGACT Symposium on Principles of Programming Languages}, pages 256--270, 2016.

\bibitem{sa4u}
Max Taylor, Johnathon Aurand, Feng Qin, Xiaorui Wang, Brandon Henry, and Xiangyu Zhang.
\newblock Sa4u: practical static analysis for unit type error detection.
\newblock In {\em Proceedings of the 37th IEEE/ACM International Conference on Automated Software Engineering}, pages 1--11, 2022.

\bibitem{avis}
Max Taylor, Haicheng Chen, Feng Qin, and Christopher Stewart.
\newblock Avis: In-situ model checking for unmanned aerial vehicles.
\newblock In {\em 2021 51st Annual IEEE/IFIP International Conference on Dependable Systems and Networks (DSN)}, pages 471--483. IEEE, 2021.

\bibitem{sel4}
{UNSW}.
\newblock sel4 secure microkernel.
\newblock \url{https://sel4.systems/}, 2013.
\newblock Accessed: 2024-06-06.

\bibitem{vassor_RMPST}
Martin Vassor and Nobuko Yoshida.
\newblock {Refinements for Multiparty Message-Passing Protocols: Specification-Agnostic Theory and Implementation}.
\newblock In Jonathan Aldrich and Guido Salvaneschi, editors, {\em 38th European Conference on Object-Oriented Programming (ECOOP 2024)}, volume 313 of {\em Leibniz International Proceedings in Informatics (LIPIcs)}, pages 41:1--41:29, Dagstuhl, Germany, 2024. Schloss Dagstuhl -- Leibniz-Zentrum f{\"u}r Informatik.

\bibitem{model1}
Nat~Andreas Wortmann.
\newblock {\em Model-Driven Architecture and Behavior of Cyber-Physical Systems}.
\newblock PhD thesis, RWTH Aachen University, 2021.

\bibitem{Zhang2022}
Qingzhao Zhang, Xiao Zhu, Mu~Zhang, and Z.~Morley Mao.
\newblock Automated runtime mitigation for misconfiguration vulnerabilities in industrial control systems.
\newblock In {\em ACM International Conference Proceeding Series}, pages 333--349. Association for Computing Machinery, 10 2022.

\bibitem{Zhou2020}
Fangyi Zhou, Francisco Ferreira, Raymond Hu, Rumyana Neykova, and Nobuko Yoshida.
\newblock Statically verified refinements for multiparty protocols.
\newblock In {\em OOPSLA}, 9 2020.

\bibitem{translationBasedModelChecking}
Min Zhou, Fei He, Ming Gu, and Xiaoyu Song.
\newblock Translation-based model checking for plc programs.
\newblock In {\em 2009 33rd Annual IEEE International Computer Software and Applications Conference}, volume~1, pages 553--562. IEEE, 2009.

\bibitem{fuzzing-roadmap}
Xiaogang Zhu, Sheng Wen, Seyit Camtepe, and Yang Xiang.
\newblock Fuzzing: a survey for roadmap.
\newblock {\em ACM Computing Surveys (CSUR)}, 54(11s):1--36, 2022.

\end{thebibliography}
\bibliographystyle{plain}

\appendices
\section{}

\subsection{Semantic Function $\mathcal{E}$ in \S\ref{sec: FT proofs}}

The semantic function is used to define the meaning behind executing each RPC. It must specify the valid and invalid cases for each external command defined in the syntax.
\label{app:sem-fun}
\begin{lstlisting}[style=fstar,frame=none]
let $\mathcal{E}$ (c : ExtCom) (s : state) : state =
  match c, s with
  | _, Wrong -> Wrong
  | IsFull, Sigma bay inp ->
      if None? (find_color' Empty 1 bay) 
      then Sigma bay inp else Wrong   
  | NotFull, Sigma bay inp ->
      if Some? (find_color' Empty 1 bay) 
      then Sigma bay inp else Wrong           
  | HasColor, Sigma bay inp -> 
      match inp with 
      | None -> Wrong
      | Some (RetrieveRequest color) ->
             if Some? (find_color' color 1 bay) 
             then Sigma bay inp else Wrong 
      | _ -> Wrong
  | NoColor, Sigma bay inp ->
      match inp with 
      | None -> Wrong
      | Some (RetrieveRequest color) ->
             if None? (find_color' color 1 bay) 
             then Sigma bay inp else Wrong 
      | _ -> Wrong
  | Store color, Sigma bay inp -> 
      match find_color' Empty 1 bay with  
      | Some n -> 
          Sigma (update_bay bay n color) inp 
      | _ -> Wrong     
  | Retrieve color, Sigma bay inp -> 
      match find_color' color 1 bay with 
      | Some n -> 
          Sigma (update_bay bay n Empty) inp 
      | _ -> Wrong
\end{lstlisting}

\subsection{Proof of Safety Condition in \S\ref{sec: FT proofs}}
The proof shows that for any valid state $\sigma$, $\langle spec, \sigma \rangle \not \Rightarrow_{IC} \textbf{Wrong}$. The proof uses the atomic propositions provided by the user to ensure that $spec$ can only generate valid traces.
\label{app:safety-proof}
\begin{lstlisting}[style=fstar,frame=none]
let lemma_session_good ($\sigma$ : state{$\sigma$ $\ne$ Wrong}) 
 : Lemma (ensures $\sigma$ spec $\not \Rightarrow_{IC}$ Wrong) = 
 match spec, $\sigma$ with
 | If APHasStoreRequest c1 c2, 
      Sigma bays inputs $\rightarrow$ begin 
   if eval_atomic_prop APHasStoreRequest $\sigma$
     && eval_atomic_prop APBaysFull $\sigma$ then
     assert (None? (find_color' Empty 1 bays))
   else if eval_atomic_prop APHasStoreRequest $\sigma$
     && $\lnot$(eval_atomic_prop APBaysFull $\sigma$) then
     assert (Some? (find_color' Empty 1 bays))
   else if eval_atomic_prop APHasRetrieveRequest $\sigma$ 
     && eval_atomic_prop APRequestRed $\sigma$ 
     && eval_atomic_prop APContainsRed $\sigma$ then
     assert (mem APContainsRed (label_fn $\sigma$))
   else if eval_atomic_prop APHasRetrieveRequest $\sigma$ 
     && eval_atomic_prop APRequestRed $\sigma$
     && $\lnot$(eval_atomic_prop APContainsRed $\sigma$) then
     assert ($\lnot$(mem APContainsRed (label_fn $\sigma$)))
   else if eval_atomic_prop APHasRetrieveRequest $\sigma$ 
     && eval_atomic_prop APRequestWhite $\sigma$ 
     && eval_atomic_prop APContainsWhite $\sigma$ then
     assert (mem APContainsWhite (label_fn $\sigma$))
   else if eval_atomic_prop APHasRetrieveRequest $\sigma$
     && eval_atomic_prop APRequestWhite $\sigma$ 
     && $\lnot$(eval_atomic_prop APContainsWhite $\sigma$) then
     assert ($\lnot$(mem APContainsWhite (label_fn $\sigma$)))
   else if eval_atomic_prop APHasRetrieveRequest $\sigma$ 
     && eval_atomic_prop APRequestBlue $\sigma$ 
     && eval_atomic_prop APContainsBlue $\sigma$ then
     assert (mem APContainsBlue (label_fn $\sigma$))
   else if eval_atomic_prop APHasRetrieveRequest $\sigma$ 
     && eval_atomic_prop APRequestBlue $\sigma$ 
     && $\lnot$(eval_atomic_prop APContainsBlue $\sigma$) then
      assert ($\lnot$(mem APContainsBlue (label_fn $\sigma$)))
   else ()
   end
 | _  $\rightarrow$ ()





 
\end{lstlisting}

\subsection{Decider}
A decider procedure is implemented as part of the dynamic attestation checking. The decider checks that traces are generated by $spec$
\label{app:decider}
\begin{lstlisting}[style=fstar,frame=none]
let rec is_trace 
   (t : IntCom) (tr : list ExtCom) (st : state) 
   : Tot bool (decreases (length tr)) = 
   match tr with
     | []  $\rightarrow$ true
     | com :: tr'  $\rightarrow$ 
       let cmd', term', state' = 
         eval_until_next_com st t in 
       match cmd' with
       | None  $\rightarrow$ false
       | Some com'  $\rightarrow$ 
         if None? term' then 
           com = com' && is_trace spec tr' state'
         else
           com = com' && 
           is_trace (unwrap_some term') tr' state'
\end{lstlisting}

\subsection{Example of Valid Traces}
This example shows how a trace can be validated by the dynamic attestation checker, given the current state and HMI inputs.
\label{app: valid traces}

\begin{lstlisting}[style=fstar,frame=none]
    $\textbf{RPCs:Notfull - Store red -  Hascolor -}$
    $\textbf{Retrieve blue -  Hascolor - Retrieve white -} $ 
    $\textbf{Notfull - Store red}$
 
    $\textbf{At state}$
 $\langle$ { bay1 : white; bay2 : blue; bay3 : empty; 
    bay4 : white; bay5 : red; bay6 : red; 
    bay7 : empty; bay8 : empty; bay9 : empty } 
    StoreRequest red $\rangle$.

    $\textbf{With the following HMI inputs:}$
 $\langle$ { bay1 : white; bay2 : blue; bay3 : empty; 
    bay4 : white; bay5 : red; bay6 : red; 
    bay7 : empty; bay8 : empty; bay9 : empty } 
    StoreRequest red $\rangle$.
 $\textbf{RPCs:Notfull - Store red}$
 
 $\langle$ { bay1 : white; bay2 : blue; bay3 : red; 
    bay4 : white; bay5 : red; bay6 : red; 
    bay7 : empty; bay8 : empty; bay9 : empty } 
    RetrieveRequest blue $\rangle$. 
 $\textbf{RPCs:Hascolor - Retrieve blue}$
 
 $\langle$ { bay1 : white; bay2 : empty; bay3 : red; 
    bay4 : white; bay5 : red; bay6 : red; 
    bay7 : empty; bay8 : empty; bay9 : empty } 
    RetrieveRequest white $\rangle$. 
 $\textbf{RPCs:Hascolor - Retrieve white}$
 
 $\langle$ { bay1 : empty; bay2 : empty; bay3 : red; 
    bay4 : white; bay5 : red; bay6 : red; 
    bay7 : empty; bay8 : empty; bay9 : empty } 
    StoreRequest red $\rangle$.
 $\textbf{RPCs:Notfull - Store red}$
 
 $\langle$ { bay1 : red; bay2 : empty; bay3 : red; 
    bay4 : white; bay5 : red; bay6 : red; 
    bay7 : empty; bay8 : empty; bay9 : empty } 
    StoreRequest red $\rangle$.
\end{lstlisting}

\subsection{Examples of Invalid Traces}
The examples listed below break the safety properties in \S \ref{sec:safety-properties}. The examples show possible $State \times ExtCom$ pairs that would trigger a fail-safe.
\label{app: Invalid traces}

\begin{lstlisting}[style=fstar,frame=none]
    $\textbf{1- Store wrong color}$
 $\langle$ { bay1 : white; bay2 : blue; bay3 : empty; 
    bay4 : white; bay5 : red; bay6 : red; 
    bay7 : empty; bay8 : empty; bay9 : empty } 
    StoreRequest red $\rangle$.
 $\textbf{RPCs: Notfull - Store blue}$

    $\textbf{2- Store with full}$
 $\langle$ { bay1 : white; bay2 : blue; bay3 : white; 
    bay4 : white; bay5 : red; bay6 : red; 
    bay7 : red; bay8 : blue; bay9 : blue } 
    StoreRequest red $\rangle$.
 $\textbf{RPCs: Notfull - Store red}$

    $\textbf{3- Command mismatch}$
 $\langle$ { bay1 : white; bay2 : blue; bay3 : empty; 
    bay4 : white; bay5 : red; bay6 : red; 
    bay7 : empty; bay8 : empty; bay9 : empty } 
    StoreRequest red $\rangle$.
 $\textbf{RPCs: Hascolor - Retrieve red}$

    $\textbf{4- Response Mismatch}$
 $\langle$ { bay1 : white; bay2 : blue; bay3 : empty;
    bay4 : white; bay5 : red; bay6 : red; 
    bay7 : empty; bay8 : empty; bay9 : empty } 
    RetrieveRequest red $\rangle$.
 $\textbf{RPCs: Notfull - Retrieve red}$

    $\textbf{5- Retrieve with no color}$
 $\langle$ { bay1 : white; bay2 : blue; bay3 : empty; 
    bay4 : white; bay5 : blue; bay6 : blue; 
    bay7 : empty; bay8 : empty; bay9 : empty } 
    RetrieveRequest red $\rangle$.
 $\textbf{RPCs: Hascolor - Retrieve red}$

\end{lstlisting}










\end{document}